\documentstyle[epsfig,galley]{mn}
\newif\ifAMStwofonts

\ifoldfss

  \newcommand{\hel}[2] {He\,{\sc #1}~$\lambda$#2}
  
\def\lesssim{\mathrel{\hbox{\rlap{\hbox{\lower4pt\hbox{$\sim$}}}\hbox{$<$}}}}
\let\la=\lesssim
\def\gtrsim{\mathrel{\hbox{\rlap{\hbox{\lower4pt\hbox{$\sim$}}}\hbox{$>$}}}}
\let\ga=\gtrsim

  \ifCUPmtlplainloaded \else
    \NewTextAlphabet{textbfit} {cmbxti10} {}
    \NewTextAlphabet{textbfss} {cmssbx10} {}
    \NewMathAlphabet{mathbfit} {cmbxti10} {} 
    \NewMathAlphabet{mathbfss} {cmssbx10} {} 
  \fi
  \ifAMStwofonts
    \ifCUPmtlplainloaded \else
      \NewSymbolFont{upmath} {eurm10}
      \NewSymbolFont{AMSa} {msam10}
      \NewMathSymbol{\upi}     {0}{upmath}{19}
      \NewMathSymbol{\umu}     {0}{upmath}{16}
      \NewMathSymbol{\upartial}{0}{upmath}{40}
      \NewMathSymbol{\leqslant}{3}{AMSa}{36}
      \NewMathSymbol{\geqslant}{3}{AMSa}{3E}

    \fi
  \fi
\fi 

\ifnfssone
  \newmathalphabet{\mathit}
  \addtoversion{normal}{\mathit}{cmr}{m}{it}
  \addtoversion{bold}{\mathit}{cmr}{bx}{it}

  \newcommand{\hel}[2] {He\,{\sc #1}~$\lambda$#2}

  \newmathalphabet{\mathbfit} 
  \addtoversion{normal}{\mathbfit}{cmr}{bx}{it}
  \addtoversion{bold}{\mathbfit}{cmr}{bx}{it}
  \newmathalphabet{\mathbfss} 
  \addtoversion{normal}{\mathbfss}{cmss}{bx}{n}
  \addtoversion{bold}{\mathbfss}{cmss}{bx}{n}
  \ifAMStwofonts
    \ifCUPmtlplainloaded \else
      \UseAMStwoboldmath
      \makeatletter
      \new@mathgroup\upmath@group
      \define@mathgroup\mv@normal\upmath@group{eur}{m}{n}
      \define@mathgroup\mv@bold\upmath@group{eur}{b}{n}
      \edef\UPM{\hexnumber\upmath@group}
      \new@mathgroup\amsa@group
      \define@mathgroup\mv@normal\amsa@group{msa}{m}{n}
      \define@mathgroup\mv@bold\amsa@group{msa}{m}{n}
      \edef\AMSa{\hexnumber\amsa@group}
      \makeatother
      \mathchardef\upi="0\UPM19
      \mathchardef\umu="0\UPM16
      \mathchardef\upartial="0\UPM40
      \mathchardef\leqslant="3\AMSa36
      \mathchardef\geqslant="3\AMSa3E
    \fi
  \fi
\fi 

\ifnfsstwo

  \newcommand{\hel}[2] {He\,{\sc #1}~$\lambda$#2}

  \DeclareMathAlphabet{\mathbfit}{OT1}{cmr}{bx}{it}
  \SetMathAlphabet\mathbfit{bold}{OT1}{cmr}{bx}{it}
  \DeclareMathAlphabet{\mathbfss}{OT1}{cmss}{bx}{n}
  \SetMathAlphabet\mathbfss{bold}{OT1}{cmss}{bx}{n}
  \ifAMStwofonts
    \ifCUPmtlplainloaded \else
      \DeclareSymbolFont{UPM}{U}{eur}{m}{n}
      \SetSymbolFont{UPM}{bold}{U}{eur}{b}{n}
      \DeclareSymbolFont{AMSa}{U}{msa}{m}{n}
      \DeclareMathSymbol{\upi}{0}{UPM}{"19}
      \DeclareMathSymbol{\umu}{0}{UPM}{"16}
      \DeclareMathSymbol{\upartial}{0}{UPM}{"40}
      \DeclareMathSymbol{\leqslant}{3}{AMSa}{"36}
      \DeclareMathSymbol{\geqslant}{3}{AMSa}{"3E}
    \fi
  \fi
\fi 

\ifCUPmtlplainloaded \else
  \ifAMStwofonts \else 
    \def\upi{\pi}
    \def\umu{\mu}
    \def\upartial{\partial}
  \fi
\fi

\title[Variable circular polarization from RR Cha]{Discovery of variable circular polarization from the
remnant of Nova Chamaeleontis 1953 (RR Cha)}
\author[P. Rodr\'\i guez-Gil \& S. B. Potter]
       {Pablo Rodr\'\i guez-Gil$^{1,2}$ and Stephen B. Potter$^3$\\
       $^1$ Department of Physics and Astronomy, University of Southampton, Southampton SO17 1BJ\\
       $^2$ Instituto de Astrof\'\i sica de Canarias, V\'\i a L\'actea, s/n. La Laguna. E-38200. Santa Cruz de Tenerife. Spain \\
       $^2$ South African Astronomical Observatory, PO
Box 9, Observatory 7935, Cape Town, South Africa}
\date{Accepted 2003.
      Received 2002}

\pagerange{\pageref{firstpage}--\pageref{lastpage}}
\pubyear{2003}

\begin{document}

\maketitle

\label{firstpage}

\begin{abstract}

We report on the discovery of variable circular polarization from the remnant
of Nova Chamaeleontis 1953 (RR Cha). The circular polarization
appears to be modulated on the primary's spin period and harmonics of the positive superhump period, with an amplitude peak-to-peak of almost $\sim
10$ per cent with both negative and positive polarization.\par

A recent study by Woudt \& Warner of RR Cha shows it to have both positive and
negative superhumps, indicating the presence of a precessing/tilted accretion disc. In addition, they also find a stable period at 1950 s, characteristic of an Intermediate Polar. RR Cha also shows deep eclipses with variable depth.\par

We propose a possible explanation for the origin of the polarized
emission. We assume that it is of cyclotron origin. The variations of the circular polarization with the proposed spin period are caused by the rotation of the compact object. The long period variability arise as a result of a precessing/tilted accretion disc, periodically obscuring either of two accretion regions on the surface of the white dwarf.\par

We also point out several similarities of RR Cha with the SW Sex stars. The lack of a time resolved spectroscopic study prevents to make any conclusion regarding its possible SW Sex nature. Hence, such spectroscopic study is encouraged.\par

RR Cha is the second old nova exhibiting variable circular polarization and we therefore encourage polarimetric observations of other nova remnants.

\end{abstract}

\begin{keywords}
accretion, accretion discs -- binaries: close -- binaries: eclipsing -- stars: individual: RR Cha -- novae, cataclysmic variables
\end{keywords}

\section{Introduction}

Cataclysmic Variables (CVs) are interacting binaries in which gas from a
late-type, main sequence star (the secondary) is being transferred to a white
dwarf (the primary), as a consequence of the secondary filling its Roche
lobe. In the absence of strongly magnetic primary, the material from the secondary spirals down to the white dwarf through an accretion disc. In Intermediate Polars (IPs) the magnetic field intensity is sufficiently high to disrupt and truncate the accretion disc
at some inner radius. Accretion is then thought to continue via accretion
curtains on to the surface of the white dwarf, where the gas is shocked to high temperatures ($\sim 10^8$ K), emitting mainly high energy and circularly polarized cyclotron radiation.

Material is almost continuously being deposited on to the surface
of the primary. Occasionally, a critical density of accreted material is
reached, resulting in a thermonuclear runaway in the primary's hydrogen
envelope. Such events are observed as huge outbursts, with material being
expelled off the binary system (Starrfield, Sparks \& Truran 1976). This is a
nova explosion. Comprehensive
reviews on classical novae and CVs can be found in Bode \& Evans (1989) and
Warner (1995), respectively.

RR Cha is the remnant of Nova Chamaeleontis 1953, a moderately fast nova (rapid
rise to maximum brightness) which reached a photographic magnitude of 7.1 at
maximum light. Little work has been done on this old nova, probably due to its
faintness (V$ > 18$ mag) and extreme southern declination ($\delta \simeq
-82^{\mathrm{o}}$). Zwitter \& Munari (1996) reported strong \hel{ii}{4686}
emission in its optical spectrum, comparable to H$\beta$, when the object was
at $V=18.9$. A recent photometric study places it at $V=18.3$ (Woudt \& Warner
2002; hereafter WW02). A nova shell around RR Cha has been detected by Gill \& O'Brien
(1998). RR Cha is an eclipsing system with an orbital period of
$P_{\mathrm{orb}}=3.362$ h. Its optical light curve at times shows either
positive or negative superhumps at 3.466 and 3.271 h, respectively (WW02). These authors also report the presence of a stable period at 1950
s in the light curves which, in addition to the strong \hel{ii}{4686} emission,
suggests that RR Cha can be an IP.

\section[]{Observations and data reduction}

RR Cha was observed on five consecutive nights from 2002 March 21--25 using the
SAAO 1.9-m telescope with the UCT polarimeter (UCTPol; Cropper 1985). The
UCTPol was operated in its simultaneous linear and circular polarimetry and
photometry mode. White light observations (3500--9000~\AA) defined by a
RCA31034A GaAs photomultiplier tube response were obtained.

Polarized and non-polarized standard stars (Hsu \& Breger 1982) were
observed throughout the observing run in order to calculate the
position angle offsets and efficiency factors. Background sky
polarization measurements were also taken at frequent intervals
during the observations. Sky subtraction from the object measurements
would then proceed after fitting polynomials to the sky measurements.

\begin{figure}
\begin{center}
  \mbox{\epsfig{file=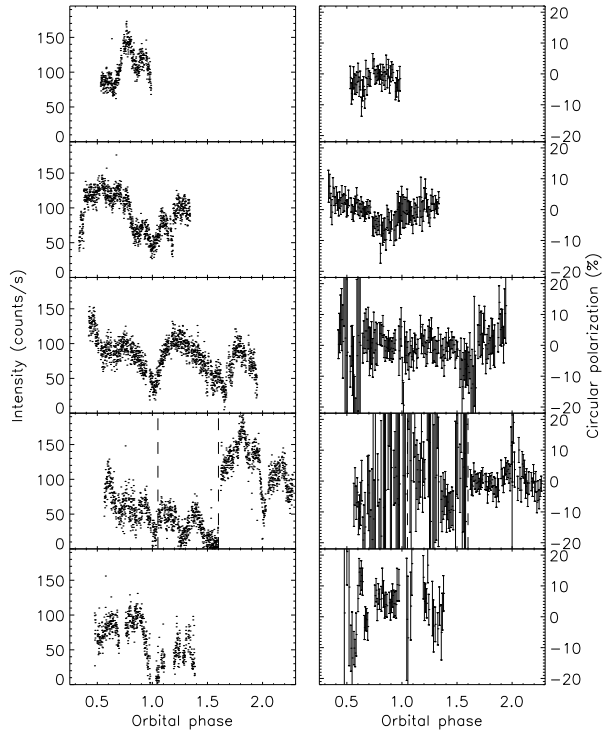,width=8cm}}
\end{center}
\caption{The photopolarimetric observations of RR Cha phased on the ephemeris
of WW02. Consecutive nights run from top to bottom. Left and
right hand panels are the photometry and circular polarization
respectively. Vertical dashed lines indicate period of non-photometric
conditions.
\label{fig:no1}}
\end{figure}

\section[]{Results}

Fig.~\ref{fig:no1} presents our photometric and circular polarization data
phased on the orbital period of WW02, and using the time of
mid-eclipse from one of our light curves. All the observations were
made under photometric conditions except for the fourth night, during which persistent thin cloud caused the loss of the guide star and consequently the partial or total disappearance of the target from the aperture. We have excluded the corresponding photometric and polarimetric data from the subsequent analysis.

\subsection[]{Photometry}

The photometric light curve displays similar characteristics to those reported in WW02. As a result of using an intensity scale, our eclipses are less
pronounced in Fig.~\ref{fig:no1} compared to the magnitude plots of WW02. There is evidence of superhump modulations, resulting in unequal
eclipse depths and morphological changes in the light curve between orbits.

\begin{figure}
\begin{center}
  \mbox{\epsfig{file=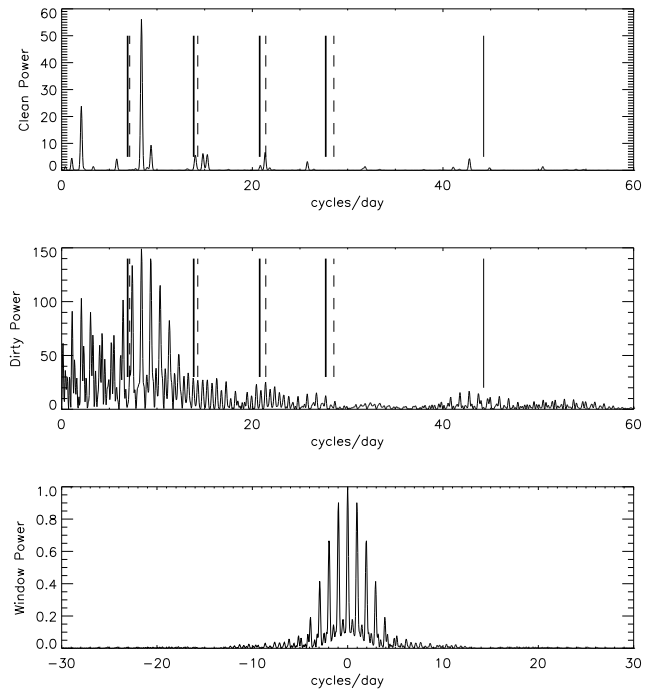,width=8.4cm}}
\end{center}
  \caption{From top to bottom: the CLEANed and the raw (dirty) power spectra, and the window function from the photometric data. Vertical lines mark the spin period (thin line), the orbital period and its harmonics (dashed lines) and the positive superhump period and its harmonics (thick lines) reported in WW02.}
\label{fig:no2}
\end{figure}

In Fig.~\ref{fig:no2} we present a Fourier analysis of the photometric data. The raw power spectrum exhibits most of the power concentrated around $\la 10$ cycles d$^{-1}$, but none of the peaks coincide with either the orbital or the superhump periods, and probably reflect the length of one or some of the data sets. The periodogram also lacks significant power at the spin frequency ($\sim 44.3$ cycles d$^{-1}$). We therefore conclude that no significant periods are detected in the photometric light curves.

\subsection[]{Circular polarimetry}

The right hand column of Fig.~\ref{fig:no1} shows the circular polarization
observations from the five consecutive nights. The first night indicated the
presence of circular polarization with a negative excursion reaching
approximately $5-10$~per cent. Follow-up observations on the second night, lasting
approximately one orbit, confirm the circularly polarized nature of RR
Cha. A sine fit to these data revealed a period of $\sim 3$ h, consistent with the findings
of WW02, and an amplitude of $\sim 5$~per cent. But the data from the third night did not produce such a clear circularly polarized signal, but both negative and positive excursions can also be seen. On the last night, the circular polarization level varied again between negative and positive values, with an amplitude of $\sim 5-10$~per cent.\par

\begin{figure}
\begin{center}
  \mbox{\epsfig{file=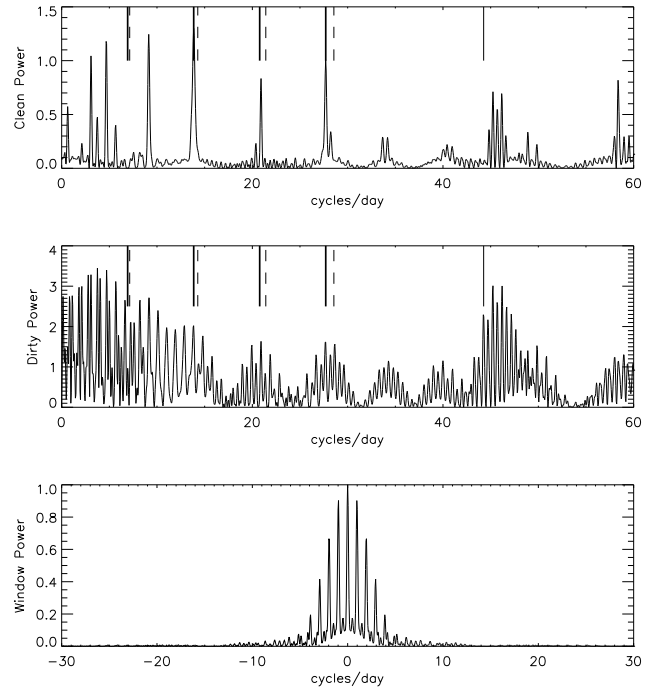,width=8.4cm}}
\end{center}
\caption{Same as Fig.~\ref{fig:no2} but from the circular polarization data.} 
\label{fig:no3}
\end{figure}

A Fourier analysis of the circular polarization curves is shown in Fig.~\ref{fig:no3}. The dirty power spectrum shows peaks coincident with the first, second, and third harmonics of the positive superhump period. The CLEANed periodogram emphasizes this more. The dirty periodogram also shows a cluster of peaks around where we expect to find the spin frequency ($\sim 44.3$ cycles d$^{-1}$). The spin period detected by WW02 coincides with one of these aliases (although it is removed by the CLEAN algorithm). All of this probably indicates that the circular polarization is modulated on the spin period and harmonics of the positive superhump period.\par

\begin{figure}
\begin{center}
  \mbox{\epsfig{file=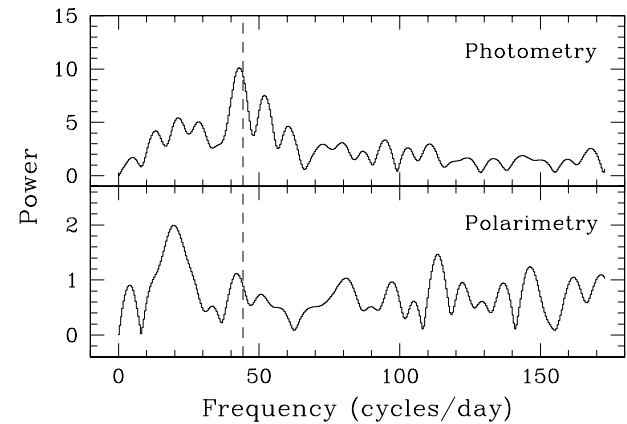,width=8.4cm}}
\end{center}
\caption{Periodograms from the sine-subtracted photometric and polarimetric curves. The dashed line marks the spin frequency reported in WW02.} 
\label{fig:no3bis}
\end{figure}

In Fig.~\ref{fig:no3bis} we construct periodograms from the photometric and polarimetric data of night 2 only, where the polarization appears more clearly defined. After subtraction of a best sine fit, the photometry shows its strongest peak very close to the assumed spin frequeny. The polarization also shows a peak at almost the same frequency, however the strongest peak is at half the spin frequency.\par

Similar separate analysis on the rest of the nights did not show such a clear spin modulation, which suggests that the regions where circularly polarized light forms were more visible on night 2. The change of behaviour between this night and, for instance, night 3 could be indicative of an obscuration of the inner disc.  

\section[]{Discussion}

The detection of significant amounts of circularly polarized light from RR Cha
is an indication of a cyclotron emission source. Alternative mechanisms such as
double scattering tend to predict smaller amounts of circularly polarized light
than observed from RR Cha. In the Polars (strongly magnetic, discless CVs) and IPs, the cyclotron emissions are generally modulated on the orbital and spin periods, respectively. In both cases, the cyclotron emission
arises from a hot shocked region near the magnetic pole on the surface of the
white dwarf.\par

RR Cha shows a modulation of the circular polarization with the probable spin period and harmonics of the superhump period. Furthermore, it displays both positive and negative circular polarization. In the synchronized Polars, this can be easily explained as cyclotron emission from two regions, each located near opposite magnetic poles. Opposite sign polarizations are also observed from IPs (e.g. PQ Gem, Potter et al. 1997). However, they give rise to a modulation on the spin period.

The photometric light curve of RR Cha presented in WW02 shows a highly coherent spin modulation at $\sim 1950$ s. The white dwarf is therefore not phase locked to the orbital motion as is the case for Polars. The photometric light curve also shows relatively wide and deep eclipses, varying eclipse depths and positive and negative superhumps. This is generally attributed to an eccentric and/or tilted precessing accretion disc.\par

\subsection[]{A modulated obscuration/accretion scenario}

The presence of a stable photometric modulation, the occurrence of wide and deep eclipses, and the detection of superhump phenomena indicate that the white dwarf in RR Cha is rotating asynchronously and possess a precessing accretion disc around it. This, together with the large circular polarization observed, confirms the IP nature of RR Cha first suggested in WW02. But the behaviour of the circular polarization curves differs greatly from those of other IPs. We naturally suggest that the precession of an eccentric/tilted disc has much to do with the observed behaviour.\par

If we closely inspect the circular polarization curves shown in Fig.~\ref{fig:no1}, we can see that the maximum excursions of the polarization are not locked in orbital phase. This of course discards a modulation at the orbital period. The polarization maxima (both positive and negative) take place at a different orbital phase each night, suggesting a modulation related to the disc precession.\par

The high inclination of RR Cha allows the precessing/tilted disc (seen nearly edge on) to periodically obscure one or the other of the cyclotron emission regions. While the disc precesses, it is possible that it can obscure our view of one of the accreting poles. Consequently, one sign of polarization is observed. At a later time, the orbit has progressed and the disc has precessed to another orientation that can obscure preferentially the other accreting pole. Therefore, the exact nature of the modulation will depend on the orbital and precession periods as well as the precession amplitudes. Moreover, the amount of material that accretes on to a particular magnetic pole would depend on the relative orientation between the magnetic field lines and the accretion disc. This will change as the white dwarf rotates and the disc precesses. Some orientations will favour accretion more than others.

\subsection[]{Similarities with the SW Sextantis stars}

The SW Sex stars are nova-like CVs which preferently show orbital periods in the range $\sim 3-4$ hours (see e.g. Mart\'\i nez-Pais, Rodr\'\i guez-Gil \& Casares 1999; Taylor, Thorstensen \& Patterson 1999; and Rodr\'\i guez-Gil \& Mart\'\i nez-Pais 2002 for an updated review of the peculiar properties of the SW Sex stars). Most of them display eclipses. They have been recently proposed as magnetic CVs based on the detection of variable circular polarization and emission-line flaring (Rodr\'\i guez-Gil et al. 2001). Their optical photometric light curves also show QPO activity with a characteristic time scale of $\ga 0.1\,P_{\mathrm{orb}}$ (see e.g. Patterson et al. 2002), resembling the well-known relation observed in IPs when the white dwarf rotation is in equilibrium with the disc. They also exhibit positive and negative superhumps in their light curves (see e.g. Rolfe, Haswell \& Patterson 2000; Patterson et al. 2002; Stanishev et al. 2002).\par

The light curve of RR Cha shows either positive or negative superhumps, deep eclipses and an orbital period of 3.36 hours, which lies on the SW Sex orbital period strip. It also displays a stable modulation (not a QPO) at $\sim 0.16\,P_{\mathrm{orb}}$. Unfortunately, no time resolved spectroscopic study has been performed on RR Cha, which could reveal a possible SW Sex nature. If RR Cha finally turns to be a SW Sex star it will provide strong evidence for the magnetic nature of the SW Sex phenomenon.

\section[]{Summary and Conclusions}

We have detected circular polarization from the remnant of Nova Chamaeleontis 1953 (RR Cha) with an amplitude peak-to-peak of $\sim 10$~per cent. We have also proposed a possible scenario in order to explain the variability of the
polarization level. It is based on the conventional picture of an IP with accretion on to two regions
close to both magnetic poles of the white dwarf. The short period modulations are due to the rotation of the primary, whereas the long period variability arises as a result of a precessing/tilted accretion disc
periodically blocking our view to either of the emission regions. There is, however, an insufficient amount of data (both photometric and polarimetric) to confirm the proposed accretion scenario.\par

RR Cha shows many similarities to the SW Sex stars. The high inclination of RR Cha is particularly favorable for time resolved spectroscopy. Such an analysis would confirm whether it is a SW Sex star or not. If the SW Sex nature is confirmed, RR Cha will be the member of the SW Sex class displaying the largest level of circular polarization.\par
 
This is the second old nova showing variable circular polarization (the first was V1500 Cyg; Stockman, Schmidt \& Lamb 1988), so further polarimetric observations of other nova remnants are encouraged.

\section{Acknowledgments}
We are grateful to the anonymous referee whose comments have improved the quality of this paper. We thank Dr. D. O'Donoghue for the use of his {\sc eagle} program for Fourier analysis of the observations, and Dr. P. Woudt and Prof. B. Warner for providing extra photometry during the polarimetry observations. We also thank Dr. E. Romero-Colmenero for comments on the original manuscript.

\bsp

\label{lastpage}

\end{document}